# STUDIES OF MATERIAL PROPERTIES UNDER IRRADIATION AT BNL LINEAR ISOTOPE PRODUCER (BLIP)*[†]

N. Simos[#], H. Kirk, H. Ludewig; BNL, Upton, NY 11973, USA
N.V. Mokhov, P. Hurh, J. Hylen, J. Misek; FNAL, Batavia, IL 60510, U.S.A.

## Abstract

Effects of proton beams irradiating materials considered for targets in high-power accelerator experiments have been under study using the Brookhaven National Laboratory's (BNL) 200 MeV Linac. The primary objectives of the study that includes a wide array of materials and alloys ranging between low and high-Z are to (a) observe changes in physio-mechanical properties which are important in maintaining high-power target functionality, (b) identify possible limits of proton flux or fluence above which certain material seize to maintain integrity, (c) study the role of material operating temperature in inducing or maintaining radiation damage reversal, and (d) correlate radiation damage effects of different species such as energetic protons and neutrons on materials by utilizing reactor and particle accelerator experience data. These objectives are specifically being addressed in the latest material irradiation study linked to the Long Baseline Neutrino Experiment (LBNE). Observations on irradiation effects on materials considered for high-power targets and other beam intercepting elements, such as collimators, from past studies and preliminary observations of the ongoing LBNE study are presented in this paper.

[*]Work supported by Fermi Research Alliance, LLC under contract No. DE-AC02-07CH11359 with the U.S. Department of Energy.
[†]Presented paper at 46th ICFA Advanced Beam Dynamics Workshop on High-Intensity and High-Brightness Hadron Beams, Sept. 27 - Oct. 1, 2010, Morschach, Switzerland.
[#]simos@bnl.gov

# STUDIES OF MATERIAL PROPERTIES UNDER IRRADIATION AT BNL LINEAR ISOTOPE PRODUCER (BLIP)*

N. Simos#, H. Kirk, H. Ludewig; BNL, Upton, NY 11973, USA
N. Mokhov, P. Hurh, J. Hylen, J. Misek; FNAL, Batavia, IL 60510, U.S.A.

*Abstract*

Effects of proton beams irradiating materials considered for targets in high-power accelerator experiments have been under study using the Brookhaven National Laboratory's (BNL) 200 MeV Linac. The primary objectives of the study that includes a wide array of materials and alloys ranging between low and high-Z are to (a) observe changes in physio-mechanical properties which are important in maintaining high-power target functionality, (b) identify possible limits of proton flux or fluence above which certain material seize to maintain integrity, (c) study the role of material operating temperature in inducing or maintaining radiation damage reversal, and (d) correlate radiation damage effects of different species such as energetic protons and neutrons on materials by utilizing reactor and particle accelerator experience data. These objectives are specifically being addressed in the latest material irradiation study linked to the Long Baseline Neutrino Experiment (LBNE). Observations on irradiation effects on materials considered for high-power targets and other beam intercepting elements, such as collimators, from past studies and preliminary observations of the ongoing LBNE study are presented in this paper.

## INTRODUCTION

High-performance targets under consideration to intercept multi-MW proton beams of a number of new particle accelerator initiatives depend almost entirely on the ability of the selected materials to withstand both the induced thermo-mechanical shock and simultaneously resist accumulated dose-induced damage which manifests itself as changes in material physio-mechanical properties. The increased demand imposed on the targets of high-power accelerators, which amounts to an order of magnitude over the experience from accelerator experiments to-date, combined with the physical limitations characterizing most common materials have led to an extensive search and experimentation with a number of new alloys and composites. In addition this search included renewed focus and interest on materials such as graphite which has been used extensively in both particle accelerators as target material and in nuclear reactors as a moderator. Driving the renewed interest in graphite is the variety of its lattice structure which may have a significant influence on its ability to operate safely at the increased demand levels of beam-induced shock and irradiation flux.

Proton irradiation effects on a wide array of materials considered to support high power experiments have been studied extensively using the BNL 200 MeV proton beam of the Linac and utilizing the target station of the Linear Isotope Producer (BLIP). Based on the Linac/BLIP parameters, and depending on the mode of operations in conjunction to the BNL accelerator complex, 20-24 kW of proton beam power (~95-100 μA current) are effectively used to irradiate target materials under consideration. While the primary goal in all the studies to-date was to both induce and subsequently observe and analyze irradiation-induced changes in the physio-mechanical properties in new alloys and composites, such as varieties of carbon-carbon, in the process, and based on experimental observations, the objective of the studies expanded to include:

(a) the identification and quantification of potentially present fluence and/or flux thresholds which may limit certain materials from operating for extended periods under MW-level operating conditions. Specifically, focus in the identification of such threshold was prompted by observations made on materials such as graphite and carbon-carbon composite which, based on reactor experience data, should have been able to maintain integrity at much higher integrated dose but appeared to be limited by a fluence threshold under proton beam irradiation,

(b) the potential role that target operating temperature may play in inducing the reversal or "healing" of radiation damage that the material undergoes due to the beam exposure. Experimental results of studies to-date using the BNL Linac beam to irradiate special alloys and composites revealed that certain lattice structures are capable of undergoing a reversal of the induced damage that is prompted by a threshold temperature which is capable of mobilizing the radiation-induced defects in the material and thus enabling the restoration of the original physical properties and

(c) the correlation of damage different irradiating species, such as energetic protons or neutrons, induce on materials as well as the energy dependence of irradiation damage. Nuclear reactor experience data on materials such as graphite exposed to primarily thermal neutrons when compared with experience data from accelerator targets where energetic protons are interacting with the same materials reveal differences in the damage rate which could be attributed to the irradiating species, the particle energy or both. Recent experimental results on graphite and carbon-carbon composites irradiated using the 200 MeV protons at BNL BLIP indicated that a threshold fluence appears to exist at ~ $0.5 \times 10^{21}$ protons/cm$^2$ beyond which these materials, which have survived much greater fluences in nuclear reactor environments, experience serious structural degradation.

___________________________________________

*Work supported by the U.S. Department of Energy.
#simos@bnl.gov

Prompted by the BNL experimental data on graphite and carbon-composites and by the observed NuMI graphite target neutron yield reduction which has been attributed to progressive target radiation damage, the interest in understanding the behavior of these materials under proton irradiation and quantifying the fluence limitations that appear to play a role has been renewed. Directly connected with the study is the Long Baseline Neutrino Experiment (LBNE) where low-Z material such as graphite, carbon-carbon composite, beryllium or its alloy AlBeMet are being considered as potential targets for the MW-level accelerator where 120 GeV protons will be intercepted. Establishing the rate of damage of the potential candidate materials of the LBNE is paramount and correlating the anticipated higher damage at the 200 MeV energies to the 120 GeV of the LBNE is the means to deducing target operating lifetime.

In the following sections results from material irradiation studies at BNL BLIP that are relevant to the high power targets and in particular to the LBNE initiative are presented and discussed. Further, details of the ongoing study directly linked to the LBNE effort, including some preliminary findings, are presented

## EXPERIMENTAL STUDIES AT BNL BLIP

Over the last decade and in an effort to identify suitable target materials for various initiatives such as muon collider/neutrino factory and the Neutrino Superbeam (currently Long Baseline Neutrino Experiment) studies using the accelerator complex at BNL have been undertaken [1]. These have in the process been augmented with studies focusing on candidate materials for the LHC collimating system as well as other accelerator components ranging from LHC calorimeter detectors, to CZT crystals and rare earth magnets for synchrotron insertion devices. With the main thrust of the effort linked to high-power accelerator targets, beam-induced shock and radiation damage have been the primary focus. The effect of intense proton pulses and the ensuing thermo-mechanical shock on the target material has been addressed with an early BNL study utilizing the 24 GeV AGS beam [2], [3]. In the effort to identify materials that can withstand thermal shock, attention was paid on the key physical property of thermal expansion which, along with other physical parameters, controls the level of resulting stresses in the target. Therefore, the ability of materials to exhibit low thermal expansion and, most importantly, to be able to maintain it after extensive beam exposure and irradiation damage of the lattice structure is very significant towards target longevity. Results of studies exploring materials throughout the atomic number range (from low-Z such as graphite to high-Z such as tungsten) are presented in this section while focusing primarily on the thermal expansion and the effects of radiation. It should be noted that thermal expansion can also very clearly reveal phase transformations in materials which may provide important clues regarding the onset of changes in the material as well as help establish operating temperatures. The effect of prolonged beam exposure on thermal expansion and the "shifting" of phase transformation onset are of primary interest. In the following, irradiation effects on materials representing the three regimes (low, mid and high-Z) are shown and discussed.

*Low-Z Materials Damage Studies*

In addition to graphite in its variety of grades which have been used extensively as primary targets interest has been focused on carbon-carbon composites and the low-Z alloy of Beryllium with aluminum AlBeMet as an alternative to pure Beryllium. Carbon-carbon composites, in particular, which appear in two- and three-dimensional weave structure (designated as 2D and 3D hereafter) exhibit very low thermal expansion along the carbon fiber orientation and are much stronger than graphite. These two attributes directly influencing shock absorbing capabilities led to their consideration for higher power targets than what graphite have served to-date and for beam intercept elements in the LHC collimators. BNL studies using the 24 GeV AGS beam with a tightly focused proton pulse (0.3mm x 0.9mm rms and $4 \times 10^{12}$ 24 GeV protons) confirmed the superiority of carbon composite structures in mitigating thermal shock. However, the ability of the fiber-reinforced carbon to resist radiation damage was widely untested which led to the radiation damage experiments at BNL BLIP along with a variety of graphite grades. Included in the matrix of low-Z materials the AlBeMet alloy (62% Be and 38% Al) was also considered and evaluated against pure Beryllium. Figure 1 depicts post-irradiation measurements of the coefficient of thermal expansion (CTE) for AlBeMet and Be at 550$^{\circ}$C as a function of proton fluence.

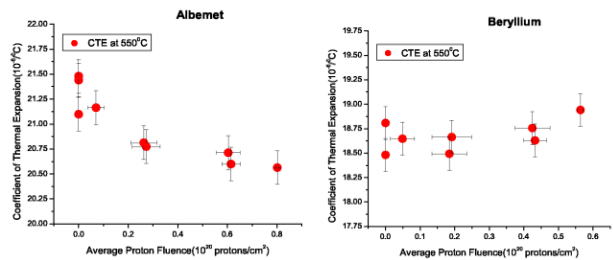

Figure 1: Thermal expansion coefficients of irradiated AlBeMet and Beryllium at elevated temperatures (550$^{\circ}$C)

Shown in Figure 2 is the effect of proton irradiation to 0.1 dpa (displacements-per-atom) on the thermal expansion of the two materials as a function of temperature. At this irradiation level the CTE of the two materials is generally unaffected.

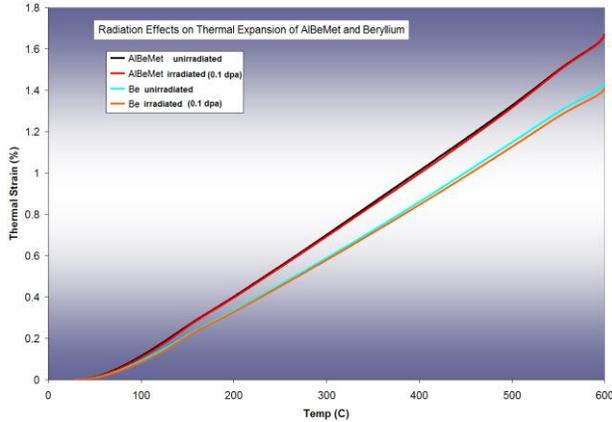

Figure 2: Comparison of the effects of proton radiation on the thermal expansion of AlBeMet and Beryllium

Figure 3 depicts a comparison of proton irradiation effects on graphite (IG-430) and the two-dimensional carbon composite studied for the LHC collimator. Shown are the initial (TC1) and final stages (TC3) of cyclic thermal annealing of the irradiated specimens (~0.2 dpa). While graphite is generally unaffected by radiation and therefore the restoration due to annealing is subtle, the CC-2D, which in the fiber plane exhibits negative thermal expansion for temperatures below ~1000°C, undergoes a significant restoration with repeated thermal cycling. Also shown in Figure 4 is the dependence of the CTE of graphite IG-43 on proton fluence observed at 550°C.

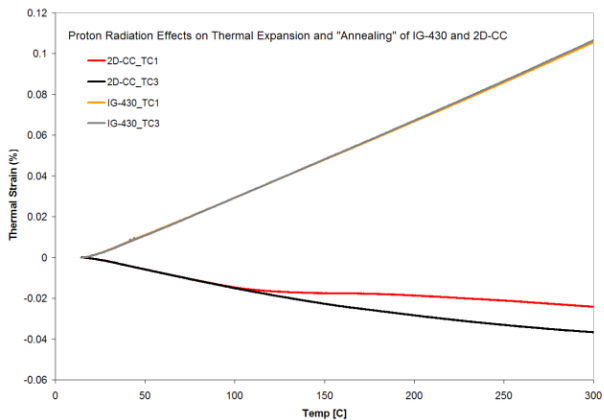

Figure 3: Proton radiation effects on carbon composite (2D structure) and graphite IG-430 CTE

To better understand the damage "annealing" properties of fiber reinforced carbon composites detailed studies where conducted comparing 2D and 3D composite structures. As shown in Figure 5 thermal cycling to a given peak temperature appears to restore part of the damage which manifests itself in the form of a dramatic change in the material thermal expansion. As shown, up to the irradiating temperature (~130°C) for this experiment) the carbon composite material is unaffected in that damage reversal takes place while the material is being irradiated. Figure 6 shows a comparison of the thermal expansion parallel the fiber plane of irradiated 2D and 3D carbon composites irradiated to a fluence of ~0.8x10$^{21}$ protons/cm$^2$ following progressive thermal cycling and damage reversal at lower temperatures. The unique behavior of these carbon structures which was observed in all the irradiation phases which led to different peak fluences was considered as a primary drive to design long-lasting high-power targets operating at temperatures where self-annealing can take place.

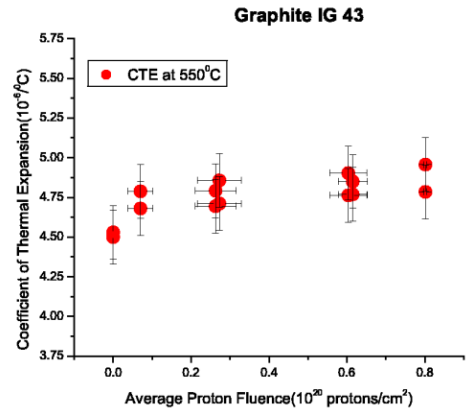

Figure 4: Thermal expansion coefficients of irradiated graphite IG-430 at 550°C.

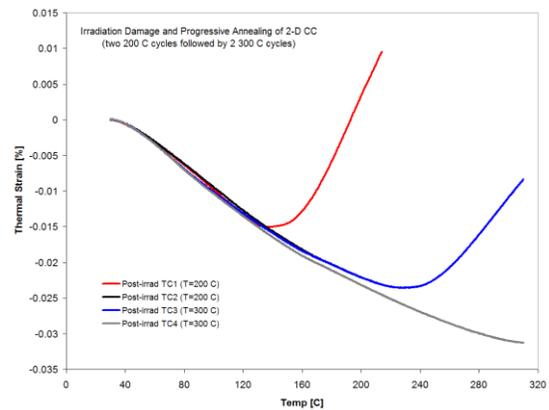

Figure 5: Progressive damage reversal through thermal cycling at increasing peak temperatures exhibited by carbon composite (2D structure) along the fiber plane

The irradiation studies which led to peak fluences > 0.5x10$^{21}$ protons/cm$^2$ also revealed that both carbon fiber structures and graphite are experiencing an accelerated structural degradation. Especially for graphite which survived in nuclear reactor core environments receiving much higher doses, this was an unexpected finding. Repeated BNL BLIP experiments under the same conditions (water-cooled target materials irradiated in the range of 120-180 MeV protons) with fluences that crossed the 0.5x10$^{21}$ protons/cm$^2$ threshold confirmed the original observations. Shown in Figures 7 and 8 are irradiated graphite specimens (IG-43) which experienced serious structural damage. Figure 7 depicts a special specimen where fusion bonding of graphite is achieved with titanium alloy Ti6Al4V. The special interface shown in the SEM graph prior to irradiation has been lost due to graphite disintegration following irradiation (insert). More clearly visible is the damage of graphite specimens exposed to the same irradiating beam.

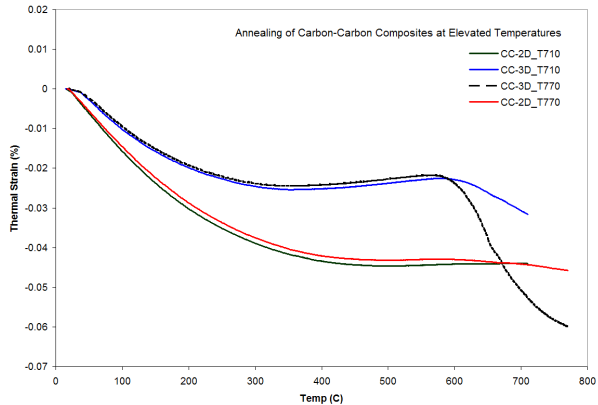

Figure 6: Annealing of irradiated 3D- and 2D-carbon composites at higher temperatures following progressive thermal cycling and annealing from 200 $^o$C to 650 $^o$C.

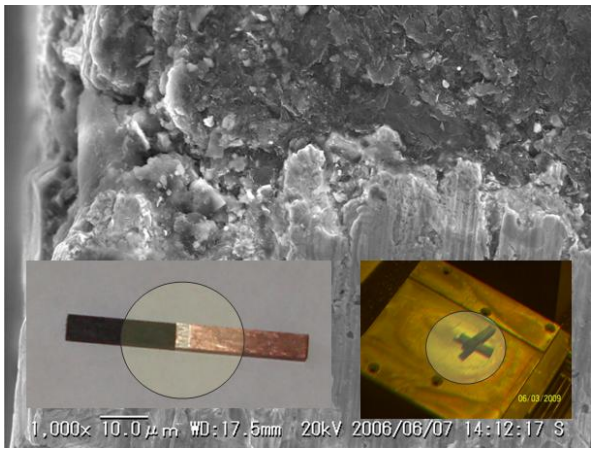

Figure 7: Observed graphite (IG-43) damage in the proximity of a special bonding interface with Ti6Al4V following irradiation at BNL BLIP

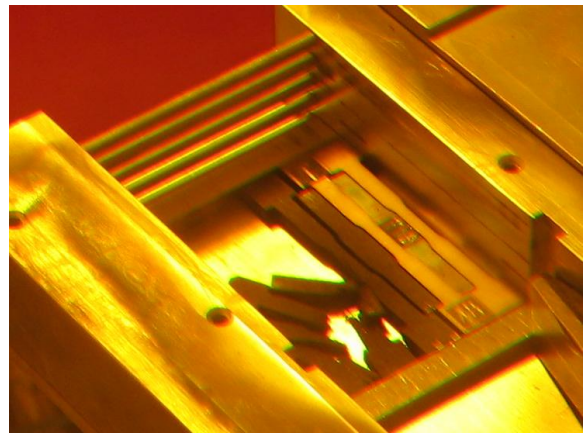

Figure 8: Observed graphite (IG-43) damage following irradiation of ~0.8x10$^{21}$ protons/cm$^2$ at BNL BLIP

*Mid-Z Range Materials Damage Studies*

Materials in this mid-Z range explored for either high power targets or beam collimating elements include super-Invar, the gum super-alloy (Ti-12Ta-9Nb-3V-6Zr-O), the titanium alloy Ti-6Al-4V, inconel-718, Cu, and Glidcop (Cu alloyed with 0.15% AlO$_3$). Target studies in this range were prompted by the extremely low thermal expansion coefficient of super-Invar up to 150$^o$C in order to help mitigate thermal shock. Following irradiation at BNL BLIP to ~0.20 dpa, it was revealed, as shown in Figure 9 that it undergoes significant change. However, and following post-irradiation annealing the threshold temperature which is required to fully restore the material was established (T$_{anneal}$ ≥ 600$^o$C). Subsequent irradiation cycles to even higher fluences revealed that while the material undergoes degradation in terms of its thermal expansion behavior, annealing above the established threshold leads to full restoration as shown in Figure 10.

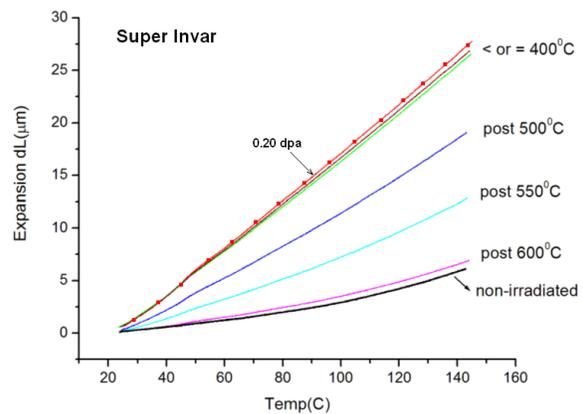

Figure 9: super-Invar post-irradiation damage reversal

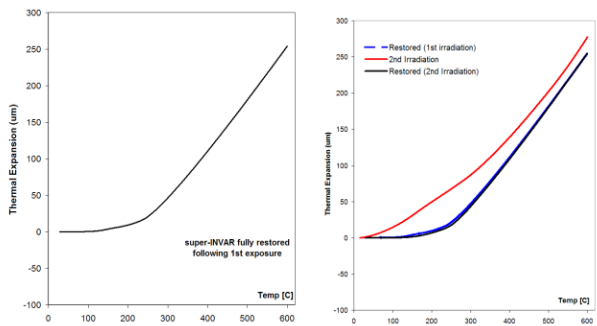

Figure10: Irradiation effects on the thermal expansion coefficient of super Invar.

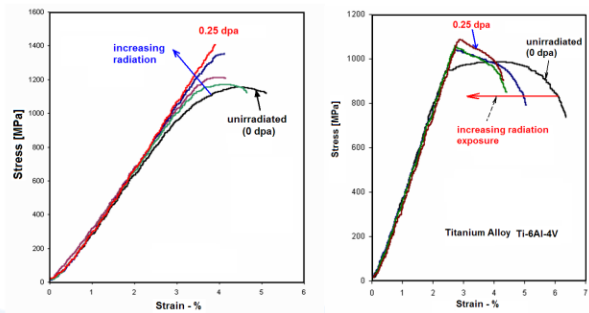

Figure12: Irradiation effects on ductility of gum metal and Ti-6Al-4V alloys

## High-Z Materials Damage Studies

In evaluating high-Z materials for use as high-power accelerator targets, tungsten and tantalum were irradiated at BNL BLIP to ~1.5 dpa. Combined with the radiation effects was the temperature and operating environment influence on these two materials. As shown in Figure 13 tantalum undergoes a phase transition at ~ 600°C which is moderately affected by radiation in that the transition temperature is lowered. A subtle phase transition (which has also been observed by other researchers in un-irradiated tungsten) at ~350°C is shown in Figure 14. The CTE of tungsten is shown to be largely unaffected by irradiation.

What is significant, however, is the loss of material due to surface oxidation of tantalum. Shown in Figure 15 is the disintegration of tantalum above 1100°C in air. The BNL study, following confirmation of the rate of oxidation observed in previous studies [4] for un-irradiated tantalum, revealed that irradiation accelerates the process.

Irradiation studies on the gum super-alloy [3], which possesses invar properties at even greater range (up to ~ 400°C) and exhibits non-elasticity with low Young's modulus and super-plastic behavior, revealed that the alloy is greatly affected by temperatures that exceed the phase transformation regime of 450-500°C and irradiation which removes its super-plastic behavior. Figure 11 depicts the thermal expansion of gum metal through the phase transformation temperature regime and shows that the transition, while unaffected by radiation, is controlled by temperature. Figure 12 depicts the dramatic loss of ductility in this alloy following modest irradiation levels and compared with the response of the Ti6Al4V alloy.

Radiation damage studies on Inconel-718 alloy, Cu and Glidcop considered for LHC Phase II collimating elements revealed that the effects on thermal expansion which is primary consideration for the intended function are very small. Assessment on thermal conductivity and stress-strain behavior is in progress.

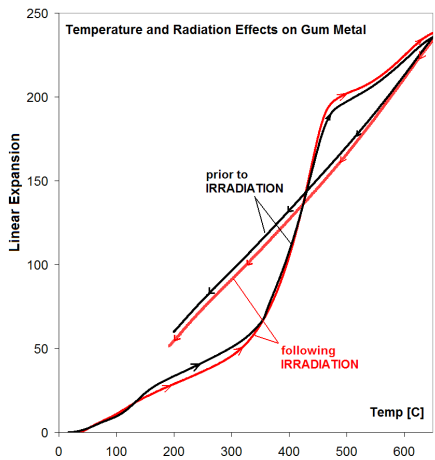

Figure11: Irradiation and temperature effects on the thermal expansion coefficient of gum metal.

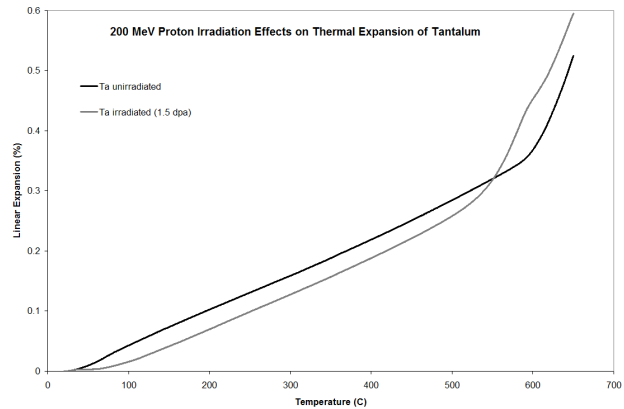

Figure 13: Irradiation effects on tantalum CTE

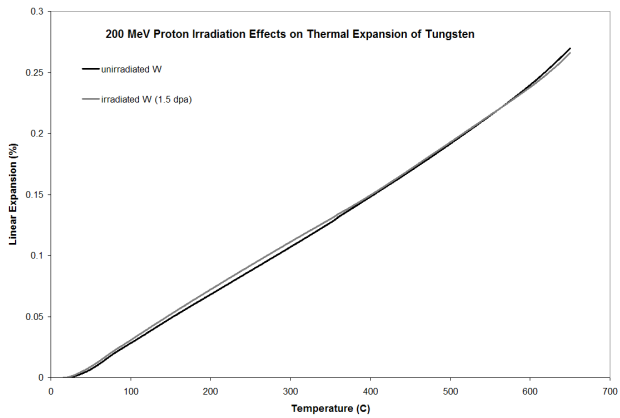

Figure 14: Irradiation effects on tungsten CTE

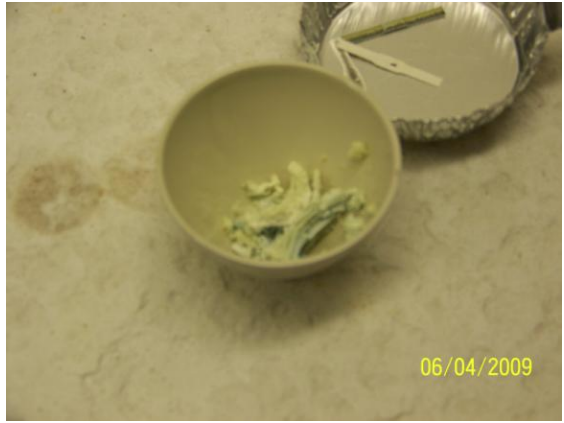

Figure 15: Oxidation of irradiated tantalum at 1100°C

## LBNE IRRADIATION DAMAGE STUDIES

The Long Baseline Neutrino Experiment is aiming to achieve a power of 2 MW where a tightly focused beam of 120 GeV protons from the Fermilab Main Injector will be intercepted by a target configuration that resembles the currently operating, but much lower power, NuMI target shown in Figure 16. The NuMI target design consists of an encapsulated and segmented rod inserted into the magnetic horn which catches and guides the liberated pions towards the decay tunnel where they decay into muons and eventually neutrinos. By implementing the same basic NuMI target design while increasing the power by approximately an order of magnitude, serious consideration to the choice of target material (currently ZXF-5Q amorphous graphite) is given. The evaluation of a number of candidate target materials in the low-Z regime that is desired for optimizing the neutrino spectra was instigated by (a) the need to identify a material that will operate at the 2 MW level with acceptable lifetime, (b) the observed degradation of yield from the NuMI target seen in Figure 17 and attributed to radiation damage in the ZXF-5Q graphite due the accumulated dose, and (c) the experimental observations at BNL on damage in graphite and carbon composites.

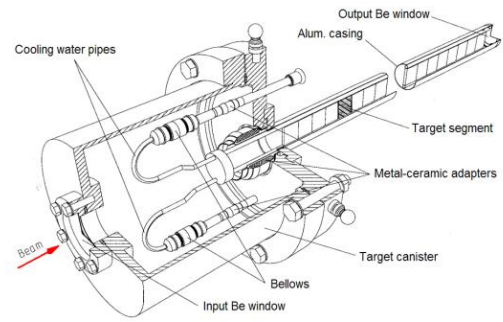

Figure 16: Operating NuMI target schematic and baseline for LBNE target concept

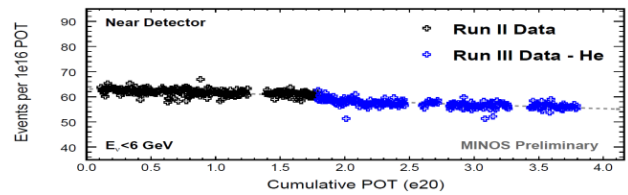

Figure 17: Observed NuMI graphite target yield degradation attributed to radiation damage

In order to qualify and quantify the feasibility of low-Z materials as target candidates for LBNE, a new experiment was conceived where the accelerated damage anticipated using the BNL BLIP facility operating to levels up to 200 MeV (as compared to the 120 GeV LBNE beam) will be utilized to evaluate a selected array. The primary goals where (a) to qualitatively assess and compare the different materials and their resilience against irradiation damage (degradation of key physio-mechanical properties) and (b) to assess the effect of operating target ambient (water-cooling vs. vacuum or inert gas) on structural integrity degradation observed in water-cooled graphite and carbon composites at BNL and a number of other accelerator experiments [5, 6].

In designing the LBNE BLIP irradiation experiment the correlation of radiation damage between the 120 GeV LBNE proton beam and the 180 MeV BNL Linac beam was sought [7]. Using the capabilities of the tracking code MARS15 [8] along with the beam operating parameters (beam spot, current, etc) the duration of the BNL BLIP experiment that is required to reveal potential damage in ~1 year of LBNE operation was established. It should be noted that radiation damage is expected to be higher in the lower energies where BLIP operates than in either the NuMI or LBNE operating with 120 GeV beam. Figures 18a and 18b show the layout of the NuMI target configuration and the baseline material matrix that was being considered for BLIP beam irradiation. The baseline array consists of several graphite grades such as ZXF-5Q POCO, Toyo-Tanso IG-430, R7650 and 2020 graphite, 3D Carbon-Carbon composite, Beryllium, AlBeMet and hexagonal Boron-Nitride.

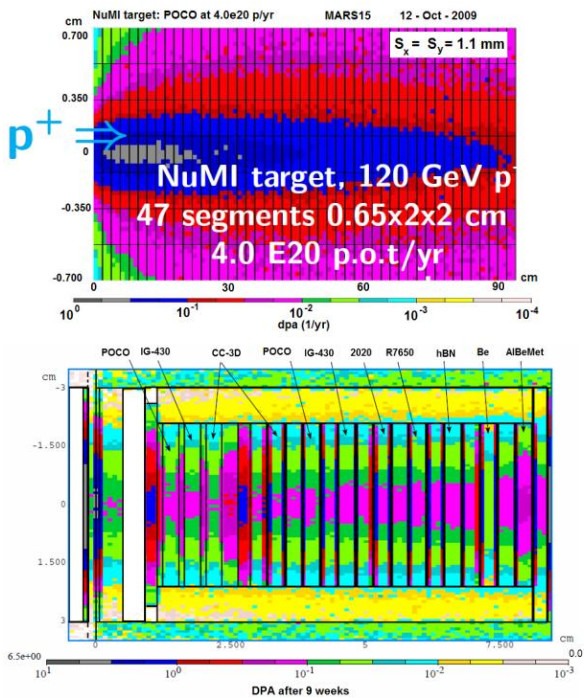

Figure 18: MARS15 comparative analysis target configurations for NuMI and the baseline array of LBNE BLIP irradiation experiment. Depicted in both model descriptions are anticipated damage in dpa.

The MARS15 analysis revealed that for the NuMI/LBNE experiment operating at 120 GeV with a beam σ = 1.1mm and 4.0e20 protons/year the expected peak damage in graphite will be 0.45 dpa while for the BLIP configuration with a beam energy of 165 MeV with beam σ = 4.23mm and 1.124e22 protons on target/year the expected damage will be 1.5 dpa. Based on these analytical results for carbon materials of interest the effect or damage of 0.7 MW LBNE operations can be achieved in ~7-8 weeks at BLIP.

To enable the isotope production at BLIP to continue uninterrupted while the LBNE targets are being irradiated modifications to the baseline matrix were made. Given that the isotope production targets which were to operate downstream of the LBNE target arrangement require a specific incoming beam energy of 112.6 MeV so the isotope yield cross-sections can be optimized, the BNL Linac operated at its 181 MeV mode. To balance the beam energy consumption through the LBNE targets while reducing the water volume in the cooling gaps such the short-lived isotopes released to the atmosphere remain below site limits, the arrangement shown in Figure 19 was finally adopted. Removed from the matrix, both for energy balance and water volume minimization are beryllium and AlBeMet and replaced with a vacuum degrader. In the configuration the upstream two layers are cooled with water to emulate the conditions of recent BLIP irradiations while the remaining six targets layers are encapsulated in a hermetically sealed argon environment. Figure 20 depicts the encapsulated arrangement of target specimens which were expected to operate at higher temperatures than the water-cooled counterparts. Detailed beam energy deposition and heat transfer studies have been performed in an effort to estimate the operating temperatures.

The LBNE BLIP irradiation study was initiated in March of 2010 and was completed in early June achieving the goal of 9-week irradiation period.

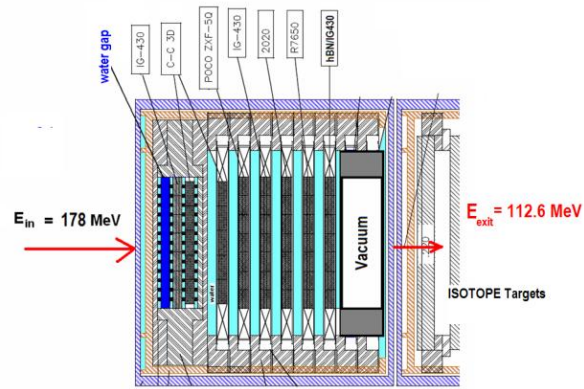

Figure 19: LBNE BLIP final target configuration

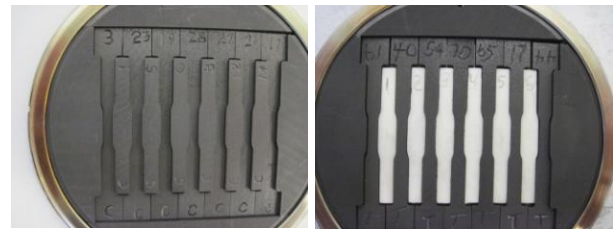

Figure 20: Specimen arrangement of encapsulated specimens in argon atmosphere

*Preliminary Results - LBNE BLIP Experiment*

Upon completion of the 9-week irradiation at BLIP with estimated 120,000 μA-hrs of beam on the LBNE targets (integrated current quoted is expected to rise by as much as 10% upon introduction of beam monitoring system correction factors) and of the "cool-down" period that allowed for sample transport to the hot cell laboratories, preliminary assessments were made on the irradiated targets.

One of the primary goals during the LBNE irradiation experiment was by exposing carbon-based materials to a fluence that exceeds the observed threshold of $0.5 \times 10^{21}$ protons/cm$^2$, which triggered serious structural damage in water-cooled graphite and carbon composites, assess the influence of the ambient environment (water vs. vacuum or inert gas) on the radiation-induced damage. Figure 21 depict the condition of the water-cooled 3D carbon composite following the 9-week irradiation. As clearly shown, structural degradation has appeared as expected given the preliminary estimates are that the threshold fluence has been exceeded during the experiment. Shown also in Figure 21 are individual 3D CC specimens. The damaged specimen in the middle of the pack is water-cooled irradiated specimen while the one at the top is one exposed to the same level but within the encapsulated argon environment and with no apparent structural damage. For reference also shown is an un-irradiated 3D CC (bottom) which appears to be in same physical condition as the argon-encapsulated irradiated specimen. This is a significant finding because it demonstrates the significant role cooling water in contact with the carbon composite surface plays in damage acceleration.

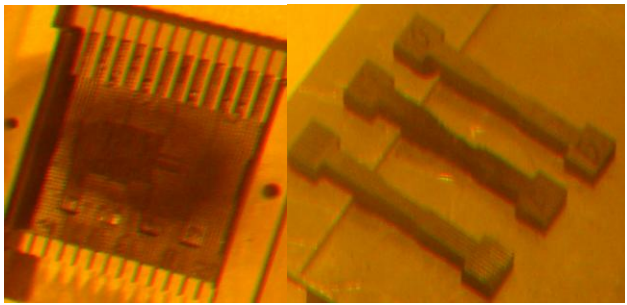

Figure 21: Irradiation damage in water-cooled 3D carbon composite LBNE targets irradiated at BLIP

To assess how irradiation, in combination with the environment, affect the physical properties of carbon composite, the thermal expansion at temperatures up to 310$^o$C were studied and compared. As shown in Figure 22, the behavior of the 3D-CC in both environments (water and argon gas) is remarkably similar to what has been observed in previous studies and the material exhibits damage reversal following thermal cycling with peak temperature greater than the irradiated temperature. Important to note in Figure 22 is the confirmation that the argon environment 3D CC specimen was operating at a higher temperature during irradiation. Following thermal cycling to 310$^o$C (temperature range is limited to avoid oxidation) both specimens reclaim the un-irradiated CTE for the regime up to 310$^o$C. More comprehensive post-irradiation analysis is under way and findings will be published in due time.

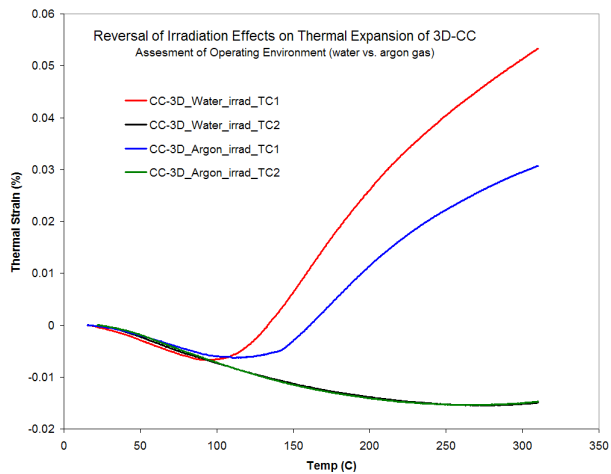

Figure 22: Thermal expansion of irradiated 3D carbon composite following irradiation and thermal cycling. Comparison between water and argon gas atmosphere.

## SUMMARY

Presented in this paper are results of an extensive irradiation study using the BNL Linear Isotope Produce facility and utilizing the 200 MeV Linac protons to irradiate a wide range of materials. Special attention has been paid to low-Z materials and in particular carbon-based (graphite and fiber reinforced carbon composites) because of (a) the numerous accelerator initiatives considering such materials for targets and beam halo intercepts, and (b) the lower-than-expected damage threshold observed when such materials are in direct contact with cooling water. Of interest also has been the assessment of irradiation-induced changes in physical properties and in particular thermal expansion of materials across the atomic number range (from low to high-Z). Experimental results and implications have been presented and discussed.

The paper discussed the most recent irradiation study at BNL linked to the LBNE targets including the objectives, and some of the preliminary results. The most important finding to-date is that lower-than-expected damage threshold in carbon-carbon composite is the direct result of the beam-water combination and not the proton beam alone. This was assessed on the basis of direct comparison of exposed carbon material under same beam parameters but different ambient conditions. This implies that by removing the environmental factor the lifetime of this material which has been also demonstrated to be superior in absorbing beam-induced shock can be extended beyond the previously thought fluence threshold.